\documentclass[12pt]{article}

\usepackage[margin=1in]{geometry}
\usepackage{amsmath, amssymb, amsthm, authblk, hyperref, CJKutf8}

\newtheorem{lemma}{Lemma}
\newtheorem{theorem}{Theorem}
\newtheorem{conjecture}{Conjecture}
\newtheorem{proposition}{Proposition}

\DeclareMathOperator{\tr}{tr}

\usepackage[style=trad-unsrt, citestyle=numeric-comp, giveninits=true, doi=false, url=false, eprint=false, isbn=false]{biblatex}
\begin{document}

\title{Entropic uncertainty relations in multidimensional position and momentum spaces}

\begin{CJK}{UTF8}{gbsn}

\author{Yichen Huang (黄溢辰)\thanks{yichuang@mit.edu} \thanks{Present address: Center for Theoretical Physics, Massachusetts Institute of Technology, Cambridge, Massachusetts 02139, USA.}}

\affil{Department of Physics, University of California, Berkeley, Berkeley, California 94720, USA}

\maketitle

\end{CJK}

\begin{abstract}

Commutator-based entropic uncertainty relations in multidimensional position and momentum spaces are derived, twofold generalizing previous entropic uncertainty relations for one-mode states. They provide optimal lower bounds and imply the multidimensional variance-based uncertainty principle. The article concludes with an open conjecture.

\end{abstract}

\section{Introduction}

Without a classical analog, uncertainty relations are one of the most fundamental ideas of quantum mechanics, underlying many conceptual differences between classical and quantum theories. They reveal by rigorous inequalities that incompatible observables cannot be measured to arbitrarily high precision simultaneously. They are applied widely in areas both related and unrelated to quantum mechanics, such as entanglement detection \cite{HHHH09, GT09, Duan, Simon, Guhne, Serafini, Hua10, Hua10E}, quantum cryptography \cite{Gisin, WW10, BB84}, and signal processing \cite{CT06, GXX09}.

We associate a random variable $A$ with an operator $\hat A$. The possible values of $A$ are the eigenvalues of $\hat A$, and the probability (density) that $A$ takes the value $a$ is the probability (density) that we get $a$ when we measure the operator $\hat{A}$ with respect to a quantum state $|\Psi\rangle$. The variance of $\hat{A}$, denoted $\Delta\hat A$, is the variance of $A$, and the (differential) Shannon entropy of $\hat A$, denoted $H(\hat A,|\Psi\rangle)$, or $H(\hat A)$, or $H(P(a))$, where $P(a)$ is the distribution of $A$, is defined as the (differential) Shannon entropy of $A$.

The famous commutator-based Heisenberg uncertainty principle is formulated by Robertson \cite{Rob29} for observables:
\begin{equation} \label{eq:h}
\Delta\hat A\Delta\hat B\ge\big|\langle\Psi|[\hat{A},\hat{B}]|\Psi\rangle\big|^2/4.
\end{equation}
We set $\hbar=1$ throughout this article. Denote the $n$-dimensional position and momentum space $H_n$. For the position and the momentum operators $\hat x,\hat p$ on $H_1$, (\ref{eq:h}) reduces to
\begin{equation} \label{eq:v}
\Delta\hat x\Delta\hat p\ge1/4.
\end{equation}

(\ref{eq:v}) is generalized to multidimensional spaces. The $n$-dimensional position and momentum space $H_n$ is described by $2n$ operators $\hat R=(\hat x_1,\hat p_1,\hat x_2,\ldots,\hat x_n,\hat p_n)$, which satisfy the canonical commutation relations $[\hat R_j,\hat R_k]=i\Omega_{jk}$, for $j,k=1,2,\ldots,2n$, where
\begin{equation}
\Omega=\bigoplus_{j=1}^n
\begin{pmatrix}
0&1\\
-1&0
\end{pmatrix}.
\end{equation}
For an $n$-mode density operator $\rho$, define the covariance matrix $\gamma$ as
\begin{equation}
\gamma_{jk}=2\tr(\rho(R_j-\tr(\rho R_j))(R_k-\tr(\rho R_k)))-i\Omega_{jk}.
\end{equation}
$\gamma$ is real and symmetric. The multidimensional variance-based uncertainty relation \cite{Simon94} is given by
\begin{equation} \label{eq:s}
\gamma+i\Omega\ge0.
\end{equation}
Reference \cite{Pirandola} provides much more detailed backgrounds. There are some other types of uncertainty relations for multimode states (e.g., Ref. \cite{Serafini}).

A different approach is to formulate uncertainty relations based on the Shannon entropy, rather than the variance. In the continuous variable case, Refs. \cite{Bec75, BM75} prove
\begin{equation} \label{eq:e}
\inf_{|\Psi\rangle}\{H(\hat x)+H(\hat p)\}=1+\ln\pi
\end{equation}
for $H_1$. Equation (\ref{eq:e}) implies (\ref{eq:v}) \cite{Bec75, BM75}, showing the advantages of entropic uncertainty relations. Lots of entropic
uncertainty relations in the discrete variable case are proposed (e.g., Ref. \cite{MU88}). Reference \cite{WW10} is a recent survey on this topic.

The main contribution of the present work is to twofold generalize Eq. (\ref{eq:e}). The commutator-based entropic uncertainty relation Eq. (\ref{eq:n}) holds for more general operators on multidimensional position and momentum spaces. Equation (\ref{eq:n}) implies the multidimensional variance-based uncertainty relation (\ref{eq:s}), so every time we use (\ref{eq:s}) in applications, we might think of using Eq. (\ref{eq:n}) instead to produce better results.

\begin{theorem} [entropic uncertainty relation in $H_n$] \label{thm:main}
\begin{equation} \label{eq:n}
\inf_{|\Psi\rangle}\{H(\hat A)+H(\hat B)\}=1+\ln\pi+\ln|[\hat A,\hat B]|,
\end{equation}
where
\begin{equation} \label{eq:f}
\hat A=\sum_{i=1}^n(a_i\hat x_i+a'_i\hat p_i),\quad\hat B=\sum_{i=1}^n(b_i\hat x_i+b'_i\hat p_i)
\end{equation}
are linear combinations of the components of $\hat R$ ($a_i,a'_i,b_i,b'_i$ are real coefficients). Equivalently and more precisely,
\begin{multline}
\inf_{|\Psi\rangle}\left\{H\left(\sum_{i=1}^n(a_i\hat x_i+a'_i\hat p_i)\right)+H\left(\sum_{i=1}^n(b_i\hat x_i+b'_i\hat p_i)\right)\right\}\\
=1+\ln\pi+\ln\left|\left[\sum_{i=1}^n(a_i\hat x_i+a'_i\hat p_i),\sum_{i=1}^n(b_i\hat x_i+b'_i\hat p_i)\right]\right|=1+\ln\pi+\ln\left|\sum_{i=1}^{n}(a_ib'_i-b_ia'_i)\right|.
\end{multline}
\end{theorem}
Letting $n=a_1=b'_1=1,a'_1=b_1=0$, Eq. (\ref{eq:n}) obviously reduces to Eq. (\ref{eq:e}). Equation (\ref{eq:n}) strengthens the importance of commutation relations and supports the intuitive idea that commutators quantify the extent of incompatibility of two operators.

The paper is organized as follows. Section \ref{s:proof} provides the detailed proof of Theorem \ref{thm:main}, which is arranged in lemmas to help you get the whole picture. Section \ref{s:dis} proves that Eq. (\ref{eq:n}) implies the variance-based uncertainty principle (\ref{eq:s}). Section \ref{s:con} concludes with an open conjecture.

\section{Proof of Theorem \ref{thm:main}} \label{s:proof}

We begin our discussion in $H_1$.

The fractional Fourier transform \cite{Namias} $\Phi(\omega)=\hat{F}(\theta)\Psi(x)$ plays an important role in the proof. It is defined as
\begin{equation}
\Phi(\omega)=\sqrt\frac{e^{i\theta-\pi i/2}}{2\pi\sin\theta}e^{\frac{i\omega^2}{2\tan\theta}}\int_{-\infty}^{\infty}e^{-\frac{i\omega x}{\sin\theta}+\frac{ix^2}{2\tan\theta}}\Psi(x)\,\mathrm dx.
\end{equation}
Naturally, $\hat F(0)$ is the identity map $\hat I$, and
\begin{equation}
\hat F(\pi/2)=\hat{\mathcal F},\quad\hat F(-\pi/2)={\hat{\mathcal F}}^{-1},
\end{equation}
where $\hat{\mathcal F}$ and $\hat{\mathcal F}^{-1}$ are the Fourier transform and the inverse Fourier transform, respectively. $\hat F$ satisfies \cite{Namias}
\begin{equation} \label{eq:g}
\hat F(\theta_1+\theta_2)=\hat F(\theta_1)\circ\hat F(\theta_2)=\hat F(\theta_2)\circ\hat F(\theta_1).
\end{equation}

The eigenvector of the operator
\begin{equation}
\hat x\cos\theta+\hat p\sin\theta=\hat x\cos\theta-i(\sin\theta)\frac{\mathrm d}{\mathrm dx}
\end{equation}
corresponding to the eigenvalue $\omega$ is
\begin{equation}
\sqrt\frac{e^{\pi i/2-i\theta}}{2\pi\sin\theta}e^{-\frac{i\omega^2}{2\tan\theta}+\frac{i\omega x}{\sin\theta}-\frac{ix^2}{2\tan\theta}}.
\end{equation}
Let $\Psi(x)$ be the position wave function of a quantum state $|\Psi\rangle$. Following from the definition of the fractional Fourier transform, the wave function in the $\hat x\cos\theta_i+\hat p\sin\theta_i$ representation is $\Psi_i=\hat F(\theta_i)\Psi(x)$ for $i=1,2$, which implies $\Psi_2=\hat F(\theta_2-\theta_1)\Psi_1$ from Eq. (\ref{eq:g}). Therefore, wave functions of the same quantum state in different representations are related by the fractional Fourier transform.

\begin{lemma} [\cite{Sha1, Sha2}] \label{l:1}
For $c\in\mathbb R$,
\begin{equation}
H(c\hat A)=H(\hat A)+\ln|c|.
\end{equation}
\end{lemma}

\begin{lemma} [\cite{GXX09}] \label{l:2}
\begin{equation} \label{eq:m}
\inf_{|\Psi\rangle}\{H(\hat{x}\cos\theta_1+\hat{p}\sin\theta_1)+H(\hat x\cos\theta_2+\hat p\sin\theta_2)\}=1+\ln\pi+\ln|\sin(\theta_2-\theta_1)|,
\end{equation}
which can be rephrased as
\begin{equation} \label{eq:l}
\inf_{\Psi(x)}\{H(|\Psi(x)|^2)+H(|\Phi(\omega)|^2)\}=1+\ln\pi+\ln|\sin\theta|,
\end{equation}
where $\Psi(x)$ runs through all legitimate wave functions and $\Phi(\omega)=\hat F(\theta)\Psi(x)$.
\end{lemma}

\begin{theorem} [Theorem \ref{thm:main} in $H_1$]
\begin{multline}
\inf_{|\Psi\rangle}\{H(a_1\hat x+a_2\hat p)+H(b_1\hat x+b_2\hat p)\}=1+\ln\pi+\ln|[a_1\hat x+a_2\hat p,b_1\hat x+b_2\hat p]|\\
=1+\ln\pi+\ln|a_1b_2-a_2b_1|.
\end{multline}
\end{theorem}

\begin{proof}
Using Lemma \ref{l:1} and then Lemma \ref{l:2}, we have
\begin{align}
&\inf_{|\Psi\rangle}\{H(a_1\hat x+a_2\hat p)+H(b_1\hat x+b_2\hat p)\}\nonumber\\
&=\inf_{|\Psi\rangle}\left\{H\left(\frac{a_1\hat x+a_2\hat p}{\sqrt{a_1^2+a_2^2}}\right)+\ln\sqrt{a_1^2+a_2^2}+H\left(\frac{b_1\hat x+b_2\hat p}{\sqrt{b_1^2+b_2^2}}\right)+\ln\sqrt{b_1^2+b_2^2}\right\}\nonumber\\
&=1+\ln\pi+\ln|a_1b_2-a_2b_1|=1+\ln\pi+\ln|[a_1\hat x+a_2\hat p,b_1\hat x+b_2\hat p]|.
\end{align}
\end{proof}

We have completed our discussion in $H_1$. Let us move on to $H_2$. Define
\begin{equation}
R_\theta=
\begin{pmatrix}
\cos\theta&-\sin\theta\\
\sin\theta&\cos\theta
\end{pmatrix}.
\end{equation}

\begin{lemma} [invariance of infimum under local rotations] \label{l:3a}
\begin{align} \label{eq:l3}
\inf_{|\Psi\rangle}\bigg\{&H\left((a_1~a_2)\begin{pmatrix}\hat x_1\\\hat p_1\end{pmatrix}+(a_3~a_4)\begin{pmatrix}\hat x_2\\\hat p_2\end{pmatrix},|\Psi\rangle\right)+H\left((b_1~b_2)\begin{pmatrix}\hat x_1\\\hat p_1\end{pmatrix}+(b_3~b_4)\begin{pmatrix}\hat x_2\\\hat p_2\end{pmatrix},|\Psi\rangle\right)\bigg\}\nonumber\\
=\inf_{|\Psi\rangle}\bigg\{&H\left((a_1~a_2)R_{\theta_1}\begin{pmatrix}\hat x_1\\\hat p_1\end{pmatrix}+(a_3~a_4)R_{\theta_2}\begin{pmatrix}\hat x_2\\\hat p_2\end{pmatrix},|\Psi\rangle\right)\nonumber\\
&+H\left((b_1\ b_2)R_{\theta_1}\begin{pmatrix}\hat x_1\\\hat p_1\end{pmatrix}+(b_3~b_4)R_{\theta_2}\begin{pmatrix}\hat x_2\\\hat p_2\end{pmatrix},|\Psi\rangle\right)\bigg\}.
\end{align}
\end{lemma}

\begin{proof}
We apply local rotations $(\hat x_1~\hat p_1)^T\to R_{\theta_1}(\hat x_1~\hat p_1)^T$ and $(\hat x_2~\hat p_2)^T\to R_{\theta_2}(\hat x_2~\hat p_2)^T$. Under this transform, a state $|\Psi\rangle$, whose position wave function is $\Psi(x_1,x_2)$, should become a new state denoted as $\hat F(\theta_1)\otimes\hat F(\theta_2)|\Psi\rangle$, whose position wave function is $(\hat F(\theta_1)\otimes\hat F(\theta_2))\Psi(x_1,x_2)$. Thus, the left-hand side of Eq. (\ref{eq:l3}) is equal to
\begin{multline}
\inf_{|\Psi\rangle}\bigg\{H\left((a_1~a_2)R_{\theta_1}\begin{pmatrix}\hat x_1\\\hat p_1\end{pmatrix}+(a_3~a_4)R_{\theta_2}\begin{pmatrix}\hat x_2\\\hat p_2\end{pmatrix},\hat F(\theta_1)\otimes\hat F(\theta_2)|\Psi\rangle\right)\\
+H\left((b_1~b_2)R_{\theta_1}\begin{pmatrix}\hat x_1\\\hat p_1\end{pmatrix}+(b_3~b_4)R_{\theta_2}\begin{pmatrix}\hat x_2\\\hat p_2\end{pmatrix},\hat F(\theta_1)\otimes\hat F(\theta_2)|\Psi\rangle\right)\bigg\}.
\end{multline}
This is equal to the right-hand side of Eq. (\ref{eq:l3}) as $\hat F(\theta_1)\otimes\hat F(\theta_2)|\Psi\rangle$ is a legitimate quantum state if and only if $|\Psi\rangle$ is a legitimate quantum state.
\end{proof}

The commutator is preserved under local rotations, which can be verified by direct computation:
\begin{multline} \label{eq:b}
\left[(a_1~a_2)\begin{pmatrix}\hat x_1\\\hat p_1\end{pmatrix}+(a_3~a_4)\begin{pmatrix}\hat x_2\\\hat p_2\end{pmatrix},(b_1~b_2)\begin{pmatrix}\hat x_1\\\hat p_1\end{pmatrix}+(b_3~b_4)\begin{pmatrix}\hat x_2\\\hat p_2\end{pmatrix}\right]\\
=\left[(a_1~a_2)R_{\theta_1}\begin{pmatrix}\hat x_1\\\hat p_1\end{pmatrix}+(a_3~a_4)R_{\theta_2}\begin{pmatrix}\hat x_2\\\hat p_2\end{pmatrix},(b_1~b_2)R_{\theta_1}\begin{pmatrix}\hat x_1\\\hat p_1\end{pmatrix}+(b_3~b_4)R_{\theta_2}\begin{pmatrix}\hat x_2\\\hat p_2\end{pmatrix}\right].
\end{multline}

\begin{lemma} [invariance of infimum under global rotations] \label{l:3b}
\begin{align} \label{eq:l4}
\inf_{|\Psi\rangle}\bigg\{&H\left((a_1~a_2)\begin{pmatrix}\hat x_1\\\hat x_2\end{pmatrix}+(a_3~a_4)\begin{pmatrix}\hat p_1\\\hat p_2\end{pmatrix},|\Psi\rangle\right)+H\left((b_1~b_2)\begin{pmatrix}\hat x_1\\\hat x_2\end{pmatrix}+(b_3~b_4)\begin{pmatrix}\hat p_1\\\hat p_2\end{pmatrix},|\Psi\rangle\right)\bigg\}\nonumber\\
=\inf_{|\Psi\rangle}\bigg\{&H\left((a_1~a_2)R_\theta\begin{pmatrix}\hat x_1\\\hat x_2\end{pmatrix}+(a_3~a_4)R_\theta\begin{pmatrix}\hat p_1\\\hat p_2\end{pmatrix},|\Psi\rangle\right)\nonumber\\
&+H\left((b_1~b_2)R_\theta\begin{pmatrix}\hat x_1\\\hat x_2\end{pmatrix}+(b_3~b_4)R_\theta\begin{pmatrix}\hat p_1\\\hat p_2\end{pmatrix},|\Psi\rangle\right)\bigg\}.
\end{align}
\end{lemma}

\begin{proof}
For simplicity, the position wave function of $|\Psi\rangle$ is denoted $\Psi((x_1~x_2)^T)$. By change of variables: $(x_1~x_2)^T\to R_\theta(x_1~x_2)^T$, which naturally yields $(\hat x_1~\hat x_2)^T\to R_\theta(\hat x_1~\hat x_2)^T$ and $(\hat p_1~\hat p_2)^T\to R_\theta(\hat p_1~\hat p_2)^T$, we see that the left-hand side of Eq. (\ref{eq:l4}) is equal to
\begin{multline}
\inf_{|\Psi\rangle}\bigg\{H\left((a_1~a_2)R_\theta\begin{pmatrix}\hat x_1\\\hat x_2\end{pmatrix}+(a_3~a_4)R_\theta\begin{pmatrix}\hat p_1\\\hat p_2\end{pmatrix},\Psi\left(R_\theta\begin{pmatrix}x_1\\x_2\end{pmatrix}\right)\right)\\
+H\left((b_1~b_2)R_\theta\begin{pmatrix}\hat x_1\\\hat x_2\end{pmatrix}+(b_3~b_4)R_\theta\begin{pmatrix}\hat p_1\\\hat p_2\end{pmatrix},\Psi\left(R_\theta\begin{pmatrix}x_1\\x_2\end{pmatrix}\right)\right)\bigg\}.
\end{multline}
This is equal to the right-hand side of Eq. (\ref{eq:l4}) as $\Psi((x_1~x_2)^T)$ is a legitimate wave function if and only if $\Psi(R_\theta(x_1~x_2)^T)$ is a legitimate wave function.
\end{proof}

The commutator is preserved under global rotations, which can be verified by direct computation:
\begin{multline} \label{eq:a}
\left[(a_1~a_2)\begin{pmatrix}\hat x_1\\\hat x_2\end{pmatrix}+(a_3~a_4)\begin{pmatrix}\hat p_1\\\hat p_2\end{pmatrix},(b_1~b_2)\begin{pmatrix}\hat x_1\\\hat x_2\end{pmatrix}+(b_3~b_4)\begin{pmatrix}\hat p_1\\\hat p_2\end{pmatrix}\right]\\
=\left[(a_1~a_2)R_\theta\begin{pmatrix}\hat x_1\\\hat x_2\end{pmatrix}+(a_3~a_4)R_\theta\begin{pmatrix}\hat p_1\\\hat p_2\end{pmatrix},(b_1~b_2)R_\theta\begin{pmatrix}\hat x_1\\\hat x_2\end{pmatrix}+(b_3~b_4)R_\theta\begin{pmatrix}\hat p_1\\\hat p_2\end{pmatrix}\right].
\end{multline}

Both local and global rotations are symplectic transformations, which preserve commutation relations (Ref. \cite{Pirandola} provides detailed relevant backgrounds). This is an alternative argument for the validity of Eqs. (\ref{eq:b}) and (\ref{eq:a}).

\begin{lemma} \label{l:4}
\begin{equation} \label{eq:o}
\inf_{|\Psi\rangle}\{H(\hat x_1)+H(\hat x_1\cos\theta+\hat p_1\sin\theta)\}=1+\ln\pi+\ln|\sin\theta|.
\end{equation}
\end{lemma}

\begin{proof}
We first show that the right-hand side is a valid lower bound and then prove its optimality. Let $\Psi(x_1,x_2)$ be the position wave function of the quantum state $|\Psi\rangle$ and $\Phi(\omega,x_2)$ be the wave function of the same state in the representation ($\hat x_1\cos\theta+\hat p_1\sin\theta,\hat x_2$). Thus, $\Phi(\omega,x_2)=(\hat F(\theta)\otimes\hat I)\Psi(x_1,x_2)$. Define
\begin{equation}
P(x_2)=\int_{-\infty}^{\infty}|\Psi(x_1,x_2)|^2\,\mathrm dx_1,
\end{equation}
which satisfies
\begin{equation}
\int_{-\infty}^{\infty}P(x_2)\,\mathrm dx_2=1.
\end{equation}
According to the definition of $H(\hat x_1)$ and due to the concavity of the Shannon entropy,
\begin{align}
H(\hat x_1)&=-\int_{-\infty}^{\infty}\left(\int_{-\infty}^{\infty}|\Psi(x_1,x_2)|^2\,\mathrm dx_2\right)\left(\ln\int_{-\infty}^{\infty}|\Psi(x_1,x_2)|^2\,\mathrm dx_2\right)\,\mathrm dx_1\nonumber\\
&=-\int_{-\infty}^{\infty}\left(\int_{-\infty}^{\infty}P(x_2)\frac{|\Psi(x_1,x_2)|^2}{P(x_2)}\,\mathrm dx_2\right)\left(\ln\int_{-\infty}^{\infty}P(x_2)\frac{|\Psi(x_1,x_2)|^2}{P(x_2)}\,\mathrm dx_2\right)\,\mathrm dx_1\nonumber\\
&\ge-\int_{-\infty}^{\infty}P(x_2)\left(\int_{-\infty}^{\infty}\frac{|\Psi(x_1,x_2)|^2}{P(x_2)}\ln\frac{|\Psi(x_1,x_2)|^2}{P(x_2)}\,\mathrm dx_1\right)\,\mathrm dx_2.
\end{align}
Similarly,
\begin{equation}
H(\hat x_1\cos\theta+\hat p_1\sin\theta)\ge-\int_{-\infty}^{\infty}P(x_2)\left(\int_{-\infty}^{\infty}\frac{|\Phi(\omega,x_2)|^2}{P(x_2)}\ln\frac{|\Phi(\omega,x_2)|^2}{P(x_2)}\,\mathrm d\omega\right)\,\mathrm dx_2,
\end{equation}
where
\begin{equation}
\Phi(\omega,x_2)=(\hat F(\theta)\otimes\hat I)\Psi(x_1,x_2)\implies\frac{\Phi(\omega,x_2)}{\sqrt{P(x_2)}}=\hat F(\theta)\frac{\Psi(x_1,x_2)}{\sqrt{P(x_2)}}.
\end{equation}
In the last equation, $\Psi$ and $\Phi$ are regarded as functions only of $x_1$ and $\omega$, respectively. Finally, applying Lemma \ref{l:2},
\begin{align}
&H(\hat x_1)+H(\hat x_1\cos\theta+\hat p_1\sin\theta)\nonumber\\
&\ge-\int_{-\infty}^{\infty}P(x_2)\left(\int_{-\infty}^{\infty}\frac{|\Psi(x_1,x_2)|^2}{P(x_2)}\ln\frac{|\Psi(x_1,x_2)|^2}{P(x_2)}\,\mathrm dx_1+\int_{-\infty}^{\infty}\frac{|\Phi(\omega,x_2)|^2}{P(x_2)}\ln\frac{|\Phi(\omega,x_2)|^2}{P(x_2)}\,\mathrm d\omega\right)\,\mathrm dx_2\nonumber\\
&\ge\int_{-\infty}^{\infty}P(x_2)(1+\ln\pi+\ln|\sin\theta|)\,\mathrm dx_2=1+\ln\pi+\ln|\sin\theta|.
\end{align}
The lower bound in Eq. (\ref{eq:l}) can be attained. Suppose it is attained for $\psi(x)$ and $\phi(\omega)$ satisfying $\phi(\omega)=\hat{F}(\theta)\psi(x)$. Let
\begin{equation}
\Psi(x_1,x_2)=\psi(x_1)\varphi(x_2),\quad\Phi(\omega,x_2)=\phi(\omega)\varphi(x_2),
\end{equation}
where $\varphi$ is an arbitrary one-dimensional legitimate wave function. In this case, it is easy to verify that the lower bound in Eq. (\ref{eq:o}) is attained, proving its optimality.
\end{proof}

\begin{lemma}
\begin{equation}
\inf_{|\Psi\rangle}\{H(\hat x_1+a\hat x_2)+H(\hat x_1\cos\theta+\hat p_1\sin\theta)\}=1+\ln\pi+\ln|\sin\theta|.
\end{equation}
\end{lemma}

\begin{proof}
Let $\Psi(x_1,x_2)$ be the position wave function of the quantum state $|\Psi\rangle$ and $\Phi(\omega,x_2)$ be the wave function of the same state in the representation ($\hat x_1\cos\theta+\hat p_1\sin\theta,\hat x_2$):
\begin{multline}
\Phi(\omega,x_2)=(\hat F(\theta)\otimes\hat I)\Psi(x_1,x_2)=\sqrt\frac{e^{i\theta-\pi i/2}}{2\pi\sin\theta}e^{\frac{i\omega^2}{2\tan\theta}}\int_{-\infty}^{\infty}e^{-\frac{i\omega x_1}{\sin\theta}+\frac{ix_1^2}{2\tan\theta}}\Psi(x_1,x_2)\,\mathrm dx_1\\
=\sqrt\frac{e^{i\theta-\pi i/2}}{2\pi\sin\theta}e^{\frac{i\omega^2}{2\tan\theta}}\int_{-\infty}^{\infty}e^{-\frac{i\omega(x_1-ax_2)}{\sin\theta}+\frac{i(x_1-ax_2)^2}{2\tan\theta}}\Psi(x_1-ax_2,x_2)\,\mathrm dx_1
\end{multline}
implies
\begin{multline}
e^{-\frac{ia\omega x_2}{\sin\theta}}\Phi(\omega,x_2)=\sqrt\frac{e^{i\theta-\pi i/2}}{2\pi\sin\theta}e^{\frac{i\omega^2}{2\tan\theta}}\int_{-\infty}^{\infty}e^{-\frac{i\omega x_1}{\sin\theta}+\frac{ix_1^2}{2\tan\theta}}e^{\frac{-2iax_1x_2+ia^2x_2^2}{2\tan\theta}}\Psi(x_1-ax_2,x_2)\,\mathrm dx_1\\
=(\hat F(\theta)\otimes\hat I)\left(e^{\frac{-2iax_1x_2+ia^2x_2^2}{2\tan\theta}}\Psi(x_1-ax_2,x_2)\right).
\end{multline}
Finally, applying Lemma \ref{l:4},
\begin{align}
&\inf_{|\Psi\rangle}\{H(\hat x_1+a\hat x_2)+H(\hat x_1\cos\theta+\hat p_1\sin\theta)\}\nonumber\\
&=\inf_{|\Psi\rangle}\left\{H\left(\int_{-\infty}^{\infty}|\Phi(\omega,x_2)|^2\,\mathrm dx_2\right)+H\left(\int_{-\infty}^{\infty}|\Psi(x_1-ax_2,x_2)|^2\,\mathrm dx_2\right)\right\}\nonumber\\
&=\inf_{|\Psi\rangle}\left\{H\left(\int_{-\infty}^{\infty}\left|e^{-\frac{ia\omega x_2}{\sin\theta}}\Phi(\omega,x_2)\right|^2\,\mathrm dx_2\right)+H\left(\int_{-\infty}^{\infty}\left|e^{\frac{-2iax_1x_2+ia^2x_2^2}{2\tan\theta}}\Psi(x_1-ax_2,x_2)\right|^2\,\mathrm dx_2\right)\right\}\nonumber\\
&=1+\ln\pi+\ln|\sin\theta|.
\end{align}
\end{proof}

Similarly,
\begin{equation} \label{eq:c}
\inf_{|\Psi\rangle}\{H(\hat x_1)+H(\hat x_1\cos\theta+\hat p_1\sin\theta+a\hat x_2)\}=1+\ln\pi+\ln|\sin\theta|.
\end{equation}

\begin{theorem} [Theorem \ref{thm:main} in $H_2$]
\begin{multline}
\inf_{|\Psi\rangle}\{H(a_1\hat x_1+a_2\hat p_1+a_3\hat x_2+a_4\hat p_2)+H(b_1\hat x_1+b_2\hat p_1+b_3\hat x_2+b_4\hat p_2)\}\\
=1+\ln\pi+\ln|[a_1\hat x_1+a_2\hat p_1+a_3\hat x_2+a_4\hat p_2,b_1\hat x_1+b_2\hat p_1+b_3\hat x_2+b_4\hat p_2]|.
\end{multline}
\end{theorem}

\begin{proof}
The idea is to reduce the most general case to more and more simpler cases by using lemmas proved previously. Assume $b_2=b_4=0$ without loss of generality. Otherwise, we apply local rotations (Lemma \ref{l:3a}), which preserve the commutator. We only need to prove
\begin{multline}
\inf_{|\Psi\rangle}\{H(a_1\hat x_1+a_2\hat p_1+a_3\hat x_2+a_4\hat p_2)+H(b_1\hat x_1+b_3\hat x_2)\}\\
=1+\ln\pi+\ln|[a_1\hat x_1+a_2\hat p_1+a_3\hat x_2+a_4\hat p_2,b_1\hat x_1+b_3\hat x_2]|.
\end{multline}
Applying global rotations (Lemma \ref{l:3b}), which preserve the commutator, we assume $b_3=0$. We only need to prove
\begin{equation}
\inf_{|\Psi\rangle}\{H(a_1\hat x_1+a_2\hat p_1+a_3\hat x_2+a_4\hat p_2)+H(b_1\hat x_1)\}=1+\ln\pi+\ln|[a_1\hat x_1+a_2\hat p_1+a_3\hat x_2+a_4\hat p_2,b_1\hat x_1]|.
\end{equation}
Assume $a_4=0$ by applying local rotations. It suffices to show
\begin{equation}
\inf_{|\Psi\rangle}\{H(a_1\hat x_1+a_2\hat p_1+a_3\hat x_2)+H(b_1\hat x_1)\}=1+\ln\pi+\ln|[a_1\hat x_1+a_2\hat p_1+a_3\hat x_2,b_1\hat x_1]|,
\end{equation}
which is equivalent to (due to Lemma \ref{l:1})
\begin{multline} \label{eq:t}
\inf_{|\Psi\rangle}\{H(\hat x_1\cos\theta+\hat p_1\sin\theta+a_3\hat x_2/a)+H(\hat x_1)\}=1+\ln\pi+\ln|[\hat x_1\cos\theta+\hat p_1\sin\theta,\hat x_1]|\\
=1+\ln\pi+\ln|\sin\theta|,
\end{multline}
where $a_1=a\cos\theta,a_2=a\sin\theta$. Now we see that Eq. (\ref{eq:t}) is precisely Eq. (\ref{eq:c}).
\end{proof}

We have completed our discussion in $H_2$. Generally, Theorem \ref{thm:main} in $H_n(n>2)$ can be proved similarly with only the following minor revision (no essentially new ideas included). We should introduce  $R\in\textnormal{SO}(n)$ ($n$-dimensional rotation) to replace the role of $R_\theta$ (two-dimensional rotation) in Lemma \ref{l:3b}, simply because the global rotation becomes an $n$-dimensional rotation in $H_n$. The $n$-dimensional version of Eq. (\ref{eq:a}) can be verified by direct computation and making use of $R\in\textnormal{SO}(n)$, or by simply using the fact that $R$ is a symplectic transformation.

We have completed the proof of Theorem \ref{thm:main}. Due to the concavity of the differential Shannon entropy, we observe that Eq. (\ref{eq:n}) also holds for mixed states, which are probabilistic mixtures of pure states.

\section{Discussions} \label{s:dis}

\begin{proposition}
Equation (\ref{eq:n}) implies the variance-based uncertainty principle (\ref{eq:s}). It also implies Serafini's multidimensional uncertainty principle---(8) in Ref. \cite{Serafini}.
\end{proposition}

\begin{proof}
We first provide an equivalent description of (\ref{eq:s}). Let $d,d'$ be two $2n$-dimensional real vectors: $d=(a_1~a'_1~a_2~\cdots~a_n~a'_n)^T$ and $d'=(b_1~b'_1~b_2~\cdots~b_n~b'_n)^T$. Obviously,
\begin{equation}
\gamma+i\Omega\ge0\iff(d+id')^\dag(\gamma+i\Omega)(d+id')\ge0~\forall d,d'.
\end{equation}
If we define operators $\hat A,\hat B$ as Eq. (\ref{eq:f}), it is equivalent to
\begin{equation}
\Delta\hat A+\Delta\hat B\ge|d'^T\Omega d|=\left|\sum_{i=1}^{n}(a_ib'_i-b_ia'_i)\right|=|[\hat{A},\hat{B}]|,
\end{equation}
because $d^T\gamma d=2\Delta\hat A$ and $d'^T\gamma d'=2\Delta\hat B$. We thus see that (\ref{eq:s}) is simply a direct consequence of the Heisenberg uncertainty principle. $H(\hat A)$ and $\Delta\hat A$ are, respectively, the differential Shannon entropy and the variance of the same distribution. From \cite{Sha1, Sha2}, we have $\Delta\hat A\ge e^{2H(\hat A)-1}/(2\pi)$. By basic inequalities, we obtain
\begin{equation}
\Delta\hat A+\Delta\hat B\ge2\sqrt{\Delta\hat A\Delta\hat B}\ge2\sqrt{\frac{e^{2H(\hat A)+2H(\hat B)-2}}{4\pi^2}}\ge\frac{e^{1+\ln\pi+\ln|[\hat{A},\hat{B}]|-1}}\pi=|[\hat{A},\hat{B}]|.
\end{equation}
Finally, Eq. (\ref{eq:n}) implies Serafini's uncertainty principle for multimode states ((8) in Ref. \cite{Serafini}) because (8) in Ref. \cite{Serafini} is a necessary condition of (\ref{eq:s}) \cite{Serafini}.
\end{proof}

\section{Conclusion and outlook} \label{s:con}

I have derived the commutator-based entropic uncertainty relation Eq. (\ref{eq:n}), which holds for more general Hermitian operators on multidimensional position and momentum spaces, twofold generalizing the previous entropic uncertainty relation Eq. (\ref{eq:e}). The lower bound in Eq. (\ref{eq:n}) is optimal, and Eq. (\ref{eq:n}) implies the multidimensional variance-based uncertainty principle (\ref{eq:s}). Every time we use (\ref{eq:s}) in applications, we might think of using Eq. (\ref{eq:n}) instead to produce better results.

One might try to seek for a simplified proof of Theorem \ref{thm:main} by using the Stone--von Neumann theorem \cite{Sto30}. However, the present proof at least has the advantage of being elementary.

A fundamental and interesting problem is to study how far we can generalize Eq. (\ref{eq:n}). We restrict $\hat A,\hat B$ to be of the form Eq. (\ref{eq:f}) in the present work, but does Eq. (\ref{eq:n}) hold for general Hermitian operators? At least we should modify Eq. (\ref{eq:n}) in the case that $[\hat{A},\hat{B}]$ is not a number operator. I propose the following open conjecture, which (if holds) implies the Heisenberg uncertainty principle Eq. (\ref{eq:h}).

\begin{conjecture}
For arbitrary Hermitian operators $\hat A,\hat B$ on multidimensional position and momentum spaces, we have
\begin{equation} \label{eq:conj}
H(\hat A,|\Psi\rangle)+H(\hat B,|\Psi\rangle)\ge1+\ln\pi+\ln\big|\langle\Psi|[\hat A,\hat B]|\Psi\rangle\big|.
\end{equation}
\end{conjecture}

If this conjecture is false, then can we add some loose restrictions on $\hat A,\hat B$ (not as strong as the restriction that $\hat A,\hat B$ should take the form of Eq. (\ref{eq:f})) so that (\ref{eq:conj}) holds?

\printbibliography

\end{document}